\documentclass[unnumsec,webpdf,contemporary,large]{oup-authoring-template}

\graphicspath{{Fig/}}

\begin{document}

\journaltitle{TBD}
\DOI{DOI to be assigned}
\copyrightyear{2025}
\pubyear{2025}
\access{Advance Access Publication Date: Day Month Year}
\appnotes{Applications Note}

\firstpage{1}

\title[DataMap: Browser-Based Data Visualization]{DataMap: A Portable Application for Visualizing High-Dimensional Data}

\author[1,$\ast$]{Xijin Ge\ORCID{0000-0001-7406-3782}}

\authormark{Ge}

\address[1]{\orgdiv{Department of Mathematics and Statistics}, \orgname{South Dakota State University}, \orgaddress{\state{South Dakota}, \country{USA}}}

\corresp[$\ast$]{Corresponding author. \href{mailto:Xijin.Ge@sdstate.edu}{Xijin.Ge@sdstate.edu}}


\abstract{
\textbf{Motivation:} The visualization and analysis of high-dimensional data are essential in biomedical research. There is a need for secure, scalable, and reproducible tools to facilitate data exploration and interpretation.\\
\textbf{Results:} We introduce DataMap, a browser-based application for visualization of high-dimensional data using heatmaps, principal component analysis (PCA), and t-distributed stochastic neighbor embedding (t-SNE). DataMap runs in the web browser, ensuring data privacy while eliminating the need for installation or a server. The application has an intuitive user interface for data transformation, annotation, and generation of reproducible R code.\\
\textbf{Availability and Implementation:} Freely available as a GitHub page \url{https://gexijin.github.io/datamap/}. The source code can be found at \url{https://github.com/gexijin/datamap}, and can also be installed as an R package.\\
\textbf{Contact:} \href{Xijin.Ge@sdstate.edu}{Xijin.Ge@sdstate.edu}
}


\maketitle

\section{Introduction}

High-dimensional datasets, such as expression matrices from RNA-seq or proteomics experiments, are routinely generated in biomedical research. Several web-based visualization tools have been developed to make these expansive datasets more accessible, including Clustergrammer \cite{fernandez2017clustergrammer}, Phantasus \cite{kleverov2024phantasus}, and Morpheus \cite{starruss2014morpheus}. Phantasus and Morpheus operate entirely within the user's browser, while Clustergrammer processes data on server-side infrastructure.

DataMap further enhances browser-based visualization by delivering high-quality graphics and reproducible R code. DataMap is an R/Shiny application deployed through Shinylive,  which is based on WebR, a special version of R  compiled into WebAssembly for execution in web browsers. Hosted on GitHub as a static file, this serverless design ensures that sensitive data remains secure on the user’s device while eliminating server constraints.

The platform supports a range of visualization techniques, including hierarchical clustering with heatmaps, principal component analysis (PCA), and t-distributed stochastic neighbor embedding (t-SNE)\cite{maaten2008tsne}, enabling researchers to identify biologically meaningful patterns, clusters, and relationships within complex datasets. Moreover, DataMap integrates seamlessly with user-provided row and column annotations, further enhancing the interpretability of its visual outputs. Our goal is to enable the generation of heatmaps and dimensionality reduction plots for general data matrices, including but not limited to omics datasets. To create a user-friendly application, the app recommends appropriate file parsing and data transformation settings by examining the file and data distribution. 

\section{Implementation}

DataMap is implemented as a Shiny application and compiled into WebAssembly using Shinylive, allowing entirely client-side execution within browsers. The app is hosted on GitHub Pages as static files, automatically exported from the source code using GitHub Actions and a workflow provided by Posit.  The source code is available at \url{https://github.com/gexijin/datamap}. Users can download the source code to run the app locally. 

\begin{enumerate}
\item \textbf{File Upload Module}: It supports diverse file formats including Excel, CSV, TSV, TXT, and other plain text formats, with automatic delimiter detection for proper parsing. 
\item \textbf{Data Transformation Module}: Provides preprocessing capabilities including log transformations, handling missing values, normalization, outlier capping, and feature filtering.
\item \textbf{Visualization Modules}: Generates heatmaps (Fig. 1A) via the pheatmap package \cite{kolde2019pheatmap}, PCA, and t-SNE (Fig. 1B) plots, offering high-quality, publication-ready visualizations. The pheatmap package supports dendrogram cutting to separate clusters of rows or columns for clearer visualization (see Fig. 1A).
\item \textbf{Code Generation Module}: Automatically records and generates reproducible R code for all analytical steps performed by users.
\end{enumerate}

\section{Features and Functionality}
\begin{enumerate}
\item \textbf{Secure Local Processing:}
DataMap processes all data securely within the browser, ensuring privacy and eliminating reliance on external server resources. This design also allows scalability to be unconstrained by server capacity.

\item \textbf{Smart Data Import:}
It automatically detects file formats, delimiters, and annotations, streamlining the data upload process. The app also examines the data to identify the presence of row and column names. Row annotations can be uploaded separately or included in the data matrix. Column annotations, such as experimental design factors in omics datasets, must be uploaded separately using matching column names.

\item \textbf{Comprehensive Data Transformations:}
The data transformation workflow employs statistical heuristics to recommend appropriate settings for effective visualization. Missing data can remain as is or be imputed using row-wise or column-wise mean or median values. When high skewness ($>$1) is detected and no negative values are present, the app recommends a log transformation, addressing common challenges associated with visualizing biological datasets.  The app infers matrix orientation by comparing row and column variability using Median Absolute Deviation and suggests appropriate centering or scaling. The mapping of data to colors in heatmaps is usually determined by the minimum and maximum values in the data matrix. This makes the mapping susceptible to outliers. Outliers beyond three standard deviations from the mean are capped, optimizing color ranges for visualization. Users can also filter out less variable rows. These built-in mechanisms ensure that even non-statisticians can use DataMap efficiently and robustly. 

\item \textbf{Publication-Quality Visualizations:}
DataMap utilizes R’s powerful visualization libraries to produce high-quality graphics that can be downloaded in PDF or PNG formats.

\begin{figure*}[!t]
\centering
\begin{minipage}[b]{0.45\textwidth}
    \centering
    \includegraphics[width=\textwidth]{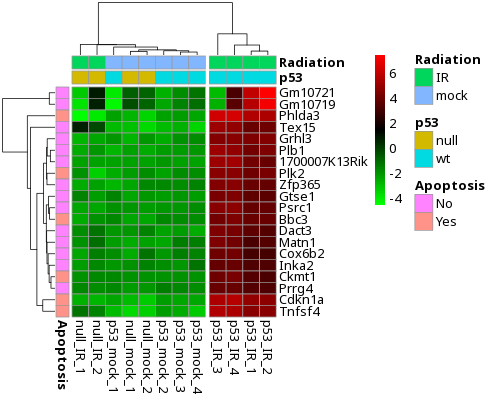}
    \\(A)
\end{minipage}
\hfill
\begin{minipage}[b]{0.4\textwidth}
    \centering
    \includegraphics[width=\textwidth]{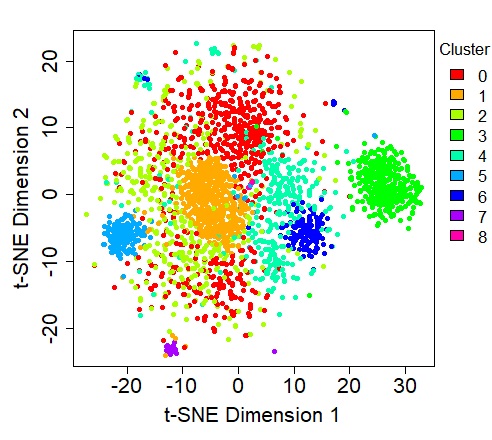}
    \\(B)
\end{minipage}
\caption{Example visualizations. (A) Top 20 genes upregulated in a p53-dependent manner by ionizing radiation in mouse B cells \cite{tonelli2015p53}, and (B) t-SNE projection of 2700 single-cell RNA-seq profiles of peripheral blood mononuclear cells (PBMCs), available from 10X Genomics. Both datasets are included as built-in examples within the application.}
\label{fig:combined}
\end{figure*}

\item \textbf{Reproducible Analysis:}
To promote transparency, consistency, and ease of collaboration, DataMap generates reproducible R code.  User settings and actions are continuously recorded to produce R code that reproduces the visualizations locally.
\end{enumerate}

\section{Comparison with Existing Tools}
DataMap complements existing visualization tools such as Clustergrammer, Phantasus, and Morpheus. Like Phantasus and Morpheus, DataMap employs client-side processing for enhanced data security. It extends their functionality by offering a broader set of preprocessing options, automatic generation of reproducible R scripts, and publication-quality graphics. However, DataMap is less interactive than native web applications built with Java or other programming languages.

\section{Discussion and Conclusion}
When analyzing large datasets, browser-based execution is slower compared to native execution. For example, generating a hierarchical clustering heatmap of a 2700×50 matrix takes approximately 80 seconds when run in the browser, compared to just 5 seconds in native R on the same laptop (Intel 11th Gen Core i7-1185G7, 3.00 GHz). Users are encouraged to install DataMap locally as an R package for extremely large datasets. For future work, we plan to explore optimization methods to improve efficiency. Another limitation stems from DataMap's reliance on the WebR, which only supports a subset of R packages with delayed updates.

DataMap represents an advancement in omics data visualization, combining secure client-side processing with robust data preprocessing and reproducible workflow generation. It complements and extends existing web-based tools, equipping biomedical researchers with a powerful tool for exploratory analysis and dissemination of findings. Future development will focus on expanding visualization capabilities and incorporating additional analytical modules.
\section{Acknowledgments}
The author is extensively assisted by AI in coding and writing.  
Supported by NIH grants (P20GM135008, R01HG010805, R01HG013534, and R43GM153076).

\bibliographystyle{plain}  

\end{document}